# ospEDA: Orthogonal Subspace Projection for Electrodermal Activity Decomposition

Yongbin Lee, Youngsun Kong, and Ki H. Chon, *Fellow, IEEE*

*Abstract*— **Electrodermal activity (EDA) is a widely used physiological signal for assessing sympathetic nervous activity, such as arousal, stress, and pain. However, reliable decomposition into tonic and phasic components remains challenging, particularly in noisy environments and across individuals with varying signal morphologies and stimulus responses. We propose ospEDA, a novel Orthogonal Subspace Projection (OSP)–based method for EDA decomposition. The method integrates (1) tonic estimation via physiologically motivated valley detection for noise robustness; (2) phasic extraction using OSP to accommodate inter-subject variability; and (3) phasic driver estimation through non-negative least squares (NNLS) deconvolution with ridge regularization.**

**We evaluated ospEDA on five real-world datasets and one simulated EDA dataset with ground-truth components, comparing its performance against six existing methods. In simulations with a 20 dB signal-to-noise ratio (SNR), ospEDA achieved the lowest root mean square error (RMSE) for estimated tonic (0.131) and phasic (0.132) components. Under noisier conditions (10 dB SNR), it maintained superior phasic RMSE (0.293), Pearson correlation (0.782), and $R^2$ (0.979) values. Furthermore, ospEDA consistently provided the highest F1-scores (0.573, 0.617, 0.638) for sympathetic nerve activity detection across 10, 20, and 30 dB SNR levels, respectively, compared to existing methods. On the real-world datasets, ospEDA achieved a stimulus classification AUROC of 0.766 and consistently maintained strong effect sizes ($\omega^2 > 0.14$) across all five datasets.**

**Overall, ospEDA represents a promising framework for EDA decomposition, showing generally consistent performance and reliable phasic driver estimation under the varying noise conditions, with potential utility for real-world physiological monitoring applications.**

## I. INTRODUCTION

ELECTRODERMAL activity (EDA) is a widely used noninvasive physiological signal for assessing sympathetic nervous system (SNS) activity [1]. EDA reflects changes in skin conductance caused by sweat gland activity, which the SNS directly modulates. The EDA signal consists of two major components: the skin conductance level (SCL), which represents the slow-varying tonic component, and the skin conductance response (SCR), which reflects the rapid and transient changes associated with sympathetic nervous system (SNS) activity. The SCR is considered an indicator of arousal events such as external stimuli, stress, or emotional responses [2].

EDA has been explored across diverse domains; however, most applications remain research-oriented, and in clinical settings it can be used as a complementary physiological indicator rather than a standalone diagnostic tool. It should be also noted that EDA has the potential to be a reliable pain indicator for those subjects who have communication issues including infants. In the context of pain, EDA can serve as a complementary indicator for autonomic responses [3]–[6]. In affective computing, EDA provides significant information about stress, arousal, and fear, thereby supporting the development of adaptive human–computer interaction (HCI) systems [7], [8]. In military and operational contexts, EDA is used to assess cognitive load and fatigue during demanding tasks [9], [10], and more recently, to predict seizures due to oxygen toxicity in Navy divers [11]. In wearable health applications, EDA sensors are increasingly used to monitor stress, anxiety, and depression in daily life [12], [13]. Despite these wide applications, robust decomposition of EDA into tonic and phasic components remains challenging, particularly in noisy environments and across diverse signal morphologies and variable stimulus responses. To mitigate this issue, some studies have extracted skin sympathetic nerve activity (SKNA) from the electrocardiogram to detect sympathetic arousal, while others have measured electrodermal activity at multiple sites [14], [15].

The overview of representative methods for EDA decomposition is presented in Table I. Continuous Decomposition Analysis (CDA) and Discrete Decomposition Analysis (DDA), both implemented in Ledalab, utilize the Bateman function-based non-negative deconvolution [16], [17]. CDA introduced continuous phasic activity measurement with a true zero baseline, addressing the problem of overlapping SCRs [16], whereas DDA applied non-negative deconvolution in a discrete diffusion modeling framework without tonic–phasic separation [17]. cvxEDA formulates EDA decomposition as a convex optimization problem with maximum a posteriori (MAP) estimation [18]. sparsEDA was designed for fast and efficient decomposition by combining Taylor series-based SCL estimation with non-negative sparse deconvolution [19]. BayesianEDA employs the expectation-maximization (EM) framework with generalized Gaussian/Laplace priors [20]. The unified dynamic model (UDM) applies a linear time-invariant (LTI) dynamic system that unifies tonic and phasic dynamics [21].

While these methods have significantly advanced EDA

Yongbin Lee, Youngsun Kong, Ki H Chon are with the Department of Biomedical Engineering, University of Connecticut, Storrs, CT 06269, USA

*Corresponding author: Ki H. Chon (e-mail: ki.chon@uconn.edu)

TABLE I
REPRESENTATIVE METHODS FOR EDA DECOMPOSITION

| Authors (Year) | Method | Approach | Results |
|---|---|---|---|
| Benedek & Kaernbach (2010) | Continuous Decomposition Analysis (CDA) | Deconvolution method for tonic and phasic activity separation, providing continuous phasic driver. | Standard error of the mean (SEM): 0.51 $\log \mu s \cdot s$ |
| Benedek & Kaernbach (2010) | Discrete Decomposition Analysis (DDA) | Non-negative deconvolution directly estimating phasic components using discrete diffusion model. | Event detection rate: 87% (vs. 78% peak detection); RMSE = 0.019 $\mu s$ |
| Greco et al. (2016) | Convex Optimization (cvxEDA) | Convex optimization with maximum a posteriori (MAP) estimation, applying sparsity and non-negativity constraints. | Page test: p = $10^{-6}$ (cvxEDA) vs. p = $10^{-3}$ (CDA). Post hoc pairwise test: p = 0.002 (cvxEDA) vs. p = 0.01 (CDA) |
| Hernando-Gallego et al. (2018) | Sparse Deconvolution (sparsEDA) | Non-negative sparse deconvolution with least absolute shrinkage and select operator (LASSO) regularization. | MSE = -48.5 dB; Computation time (s) / Signal Duration (min) = $10^{-2}$ (vs. cvxEDA $10^{-1}$, LedaLab-CDA $10^{0}$) |
| Amin et al. (2022) | Bayesian state-space model (BayesianEDA) | Expectation maximization (EM) framework with generalized Gaussian/Laplace priors. | $R^2$>0.98<br>AUC = 0.82 (arousal vs. non-arousal) |
| Wang et al. (2025) | Unified Dynamic Model (UDM) | Linear time-invariant (LTI) dynamic system that unifies tonic and phasic dynamics, with sudomotor nerve activity estimated via an EM-based algorithm. | Significant sensitivity in differentiating high vs. low arousal (p-value < 0.001) |

research, persistent challenges include (1) noise sensitivity, (2) inter-subject variability, and (3) reliable estimation of sparse driver of sympathetic nervous activity (SNA). In addition, every evaluation has been performed on an independent single dataset, leading to (4) limited comparability and inconsistent results across studies.

To address these challenges, we propose a novel EDA decomposition algorithm, ospEDA, an orthogonal subspace projection (OSP) based method for EDA decomposition. To validate (1) noise sensitivity and (3) accuracy in SNA estimation, we test the algorithm on a simulated EDA dataset with controlled noise levels. To account for (2) inter-subject variability, we performed a bootstrap-based analysis to assess the stability of the performance estimates across individuals. Finally, to overcome (4) limited comparability, we evaluated ospEDA across five diverse real-world datasets and compared its performance against six widely used baseline methods: LedaLab-CDA [16], LedaLab-DDA [17], cvxEDA [18], sparsEDA [19], BayesianEDA [20], UDM [21].

## II. DATASETS

In this study, we evaluated our method on five different real-world datasets from three different databases. Pain modality datasets provide well-defined intensity levels that enable informative comparative analyses for benchmarking. In all datasets, stimulus intensities were determined using subject-specific calibration rather than fixed levels. Furthermore, to evaluate both continuous signal reconstruction and discrete SNA detection against a known ground truth, we generated simulated EDA signals derived from established physiological models [1], [18], [20].

### A. Simulated EDA Dataset

We generated 100 simulated EDA segments of 5 minutes each. The sparse driver impulses (SNA events) were randomly distributed based on the following rules: (1) a minimum inter-event distance of 2 seconds, (2) a total of 30 events per segment, averaging one event every 10 seconds, and (3) driver amplitudes ranging from 0.01 to 1.00 $\mu S$, ensuring all simulated events exceeded the standard minimum amplitude threshold [1]. The phasic component of the EDA was obtained by convolving the SNA event sequence with a bi-exponential impulse response function (IRF), where $\tau_1$ = 0.7s, $\tau_2$ ranged from 2 to 4 s [18]. To ensure physiologically realistic temporal dynamics, the IRF kernel was limited to a duration of 20 seconds, corresponding to a cutoff frequency of 0.05 Hz, which has been shown to yield optimal SCR extraction [22], [23]. The tonic component was generated as a slowly varying drift signal, with offset (3 to 6 $\mu S$), a linear slope (-2 to 2 $\mu S$), and a sinusoidal wave (amplitude 0.1 to 0.5 $\mu S$, period 60 to 120 s, random phase), ensuring that the tonic signal remained non-negative throughout ($\geq$ 0.5 $\mu s$). Finally, additive white gaussian noise (AWGN) was introduced to generate four different signal-to-noise (SNR) conditions (clean, 30 dB, 20 dB, and 10 dB) to evaluate our method's robustness against varying degrees of sensor noise. Fig. 1 shows a representative example of the simulated EDA signal.

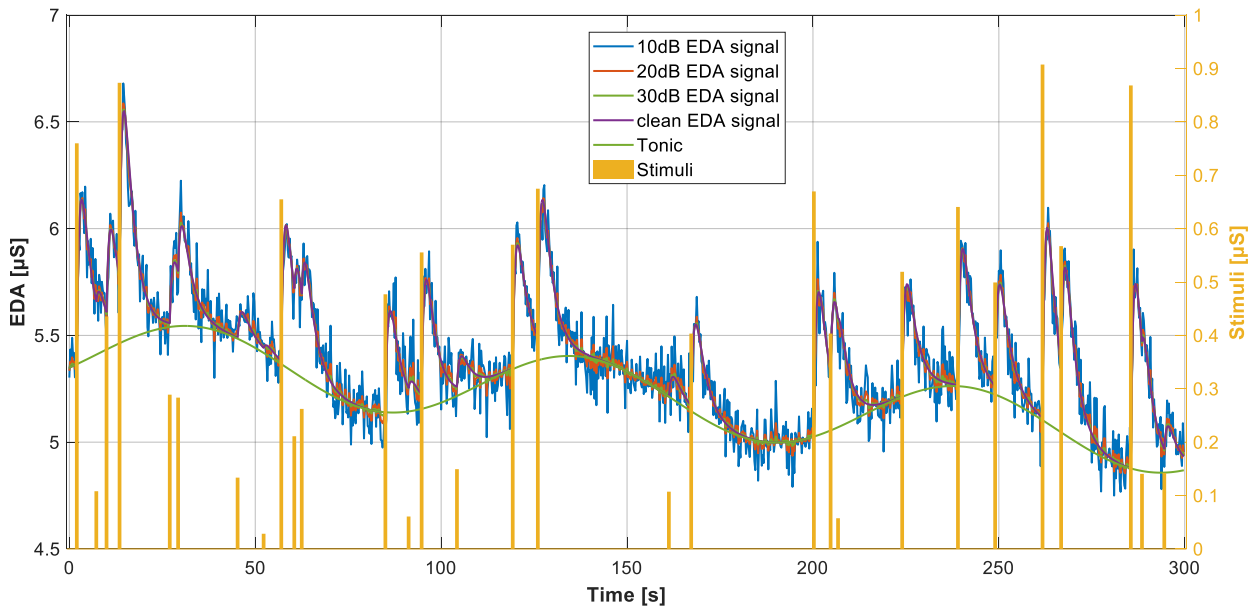

Fig. 1. An example of the simulated EDA signal. The figure displays the ground-truth tonic component, discrete sympathetic stimuli, and the composite EDA signal corrupted by varying levels of noise (clean, 30 dB, 20 dB, and 10 dB SNR).

### B. BioVid Heat Pain Database

To evaluate our EDA decomposition methods, we used the EDA signals from a total of 87 subjects in the publicly available BioVid Heat Pain Database Part C [24], [25]. The heat pain stimuli were induced using a thermode applied to the right arm, delivering five-second thermal stimulations at four different intensity levels. Four intensity levels were individually calibrated for each subject: the pain threshold (first sensation of stinging), the pain tolerance threshold (maximum tolerable heat), and two intermediate levels corresponding to the 33% and 67% points between threshold and tolerance. Each temperature level was presented 20 times, resulting in 80 stimuli per subject. Randomized recovery intervals of 8–12 seconds were inserted between stimuli. Since the stimulus labels were originally annotated without accounting for latency, they were shifted forward by 3 seconds to match the timing of the other datasets, as described in a prior study [26].

### *C. ChonLab Pain Stimulation Database*

The ChonLab Pain Database includes two experimental datasets previously published in [27]. A total of 23 healthy volunteers (11 males/ 12 females, aged 19–34) participated in the experiment. 16 subjects participated in electric pulse (EP) stimulation, and 23 subjects participated in thermal grill (TG) stimulation.

1) Electric Pulses: EP stimulation was delivered to the right forearm through two disposable Ag/AgCl electrodes. The maximum intensity was individually calculated to a subject-reported pain score of 7 out of 10 on an 11-point visual analog scale (VAS). Four intensity levels were defined by 25%, 50%, 75%, and 100% of the maximum.
2) Thermal Grill: Three-level TG stimulations were delivered to the glabrous skin of the right palm. Subjects were asked to put their right hand on the grill for 5 seconds, but they were allowed to withdraw earlier. The TG consists of interlaced warm and cool aluminum or copper tubes that induce pain illusion without tissue injury. The warm temperature was individually adjusted to elicit a subject-reported pain score of 5–6 out of 10 on the VAS, defined as Level 2. Two additional levels were then established: Level 1 = Level 2 − 2°C and Level 3 = Level 2 + 2°C, with the maximum temperature limited to 47°C for safety. Each subject completed 21 randomized TG trials (7 per intensity level).

### *D. ChonLab Stroop−Pain Database*

The ChonLab Stroop−Pain Database included three experimental datasets previously published in [28]. A total of 10 healthy subjects (5 males/5 females, aged 22–34 years) participated in the experiment. The experimental protocol consisted of four periods: (1) 2-minute baseline, (2) 2-minute cognitive stress task induced by the Stroop test, (3) pain-only stimulation induced by thermal grill, and (4) simultaneous pain and Stroop task. In this study, to compare the different stimulus levels, the Stroop-only stimulation (2) was excluded.

1) Pain only stimulation: the TG was adjusted to elicit two individual pain levels: low pain (subject reported pain score of 1–6 out of 10) and high pain (pain score of 7–10 out of 10). Each subject completed three low-pain and three high-pain trials, each lasting 30–60 seconds. Subjects reported pain intensity scores after each stimulus.
2) Pain with Stroop stimulation: In the Stroop test, subjects were verbally asked to state the color of the font displayed on a screen, while the text spelled out a different color (e.g., the word “blue” written in green font). The font color changed every 1–3 seconds, requiring continuous responses. During this task, subjects simultaneously performed the pain test and the Stroop test.

## III. Methods

The EDA signal can be decomposed into two different components: the skin conductance level (tonic) and the skin conductance response (phasic). While early in vivo micropipette studies observed mechanical sweat gland activity at frequencies of 0.5–2 Hz, the electrical manifestation of these responses recorded via surface electrodes is heavily smoothed by the stratum corneum [1]. Consequently, to optimally extract the phasic component from surface recordings, the primary frequency band has been established as 0.045 to 0.25 Hz [22], [23]. However, to preserve the full frequency band of the SCRs, we applied a wider bandwidth during initial preprocessing. Therefore, the EDA signal was first low-pass filtered using an $8^{th}$-order Chebyshev Type I anti-aliasing filter (cutoff frequency = 1.6 Hz) and downsampled to 4 Hz.

### *A. Initial Tonic Estimation*

To estimate the initial tonic baseline, we must first detect the phasic-related valleys in the EDA; therefore, the downsampled signal was first quadratically detrended to remove baseline drift. Valleys were then detected according to the following criteria: (1) a minimum prominence higher than 0.05 $\mu S$, and (2) a minimum inter-valley distance of 20 seconds. The prominence threshold of 0.05 $\mu S$ was chosen because phasic responses are defined with amplitudes greater than this value, while the 20-second inter-valley spacing reflects tonic frequency ranges [22], [23]. To ensure robustness of the spline interpolation procedure as identified according to the criteria described above, two additional rules were applied. First, control points were assigned to the start and end of each segment. Second, if any 20-second window contained no detected valleys, the local minimum within that window was inserted as a control point. After detection, valley indices were remapped to the original sampling rate of the EDA signal. Finally, cubic spline interpolation across the valley points was used to estimate the tonic component, with the residual representing the phasic component.

### *B. Orthogonal Subspace Projection for Re-estimation of the Tonic Component*

This initial tonic estimate obtained by spline interpolation is influenced by the detected valley points. Therefore, it may be affected by local fluctuations caused by phasic responses, and it highly relies on the valley detection algorithm. To reduce this bias, we re-estimate the tonic component using orthogonal subspace projection.

Let $\mathrm{Y} = \{y_i\}_{i=1}^{N}$ denote the raw EDA signal and $\widehat{\mathrm{X}} = \{x_i\}_{i=1}^{N}$ the estimated tonic component as described above. To determine the optimal model order $m$, which defines the number of sample delays (lags) of the initial tonic estimate used to span the projection subspace, we constructed delayed versions of $\widehat{X}$ up to a maximum delay of a second (corresponding to a frequency of 1 Hz). For each candidate model order m, the lag matrix is defined as:

$$V_m = \begin{bmatrix} x_1 & \cdots & x_{m+1} \\ \vdots & \ddots & \vdots \\ x_{N-m} & \cdots & x_N \end{bmatrix} \quad (1)$$

where each column is a delayed version of $\widehat{X}$. Since $V_m$ has $N-m$ rows, the corresponding truncated signal is:

$$Y_m = \{y_{m+1}, y_{m+2}, \dots, y_N\} \quad (2)$$

The projection of the truncated signal is then:

$$\hat{Y}_m = P_m Y_m \quad (3)$$

where $\hat{Y}_m$ denotes the estimated signal, and $P_m$ is the projection matrix onto the subspace spanned by $V_m$, given by:

$$P_m = \begin{cases} V_m(V_m^\top V_m)^{-1}V_m^\top, & \kappa_1(V_m^\top V_m)^{-1} \geq 10^{-8} \\ V_m(V_m^\top V_m + \lambda I)^{-1}V_m^\top, & \kappa_1(V_m^\top V_m)^{-1} < 10^{-8} \end{cases} \quad (4)$$

Where the reciprocal condition number of $V_m^\top V_m$ is defined as:

$$\kappa_1(V_m^\top V_m)^{-1} = \frac{1}{\|V_m^\top V_m\|_1 \|(V_m^\top V_m)^{-1}\|_1} \quad (5)$$

Here $\kappa_1(\cdot)$ denotes the 1-norm condition number. A small reciprocal value indicates that $V_m^\top V_m$ is ill-conditioned, in which case the regularization ($\lambda = 0.01$) is used. The residual is defined as:

$$\varepsilon_m = Y_m - \hat{Y}_m \quad (6)$$

The residual variance is defined as:

$$\sigma_m^2 = \frac{1}{N}\sum_{i=1}^{N} \varepsilon_{m,i}^2 \quad (7)$$

For each order $m$, the Minimum Description Length (MDL) is computed as:

$$MDL(m) = N\ln(\sigma_m^2) + (m+1)\log(N-m) \quad (8)$$

Finally, the optimal model order $m^*$ was selected as the value of minimizing $MDL(m)$. With the selected optimal model order $m^*$, the lag matrices $V_{m^*}$ and $P_{m^*}$ are obtained Eqs. (1) – (3). The tonic component is reconstructed as:

$$Y_{tonic} = P_{m^*} Y_{m^*} \quad (9)$$

Finally, the orthogonal subspace projected phasic component is estimated as:

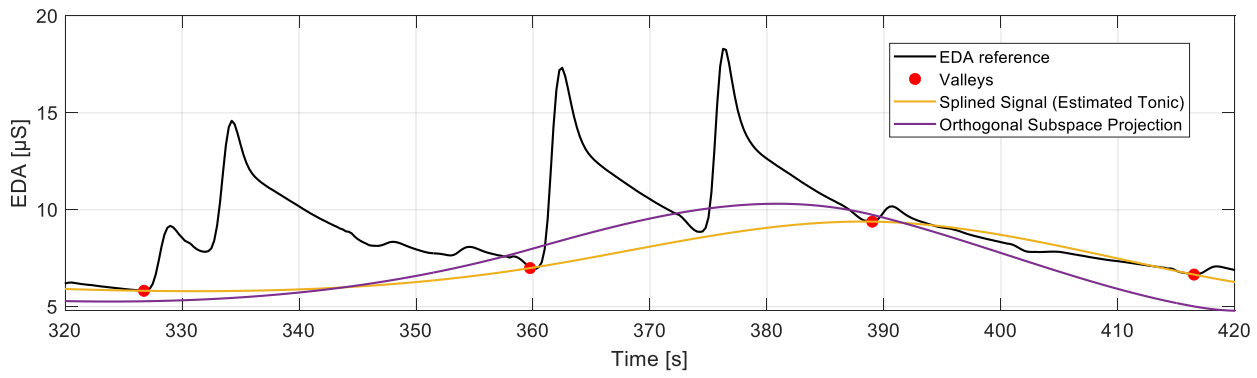

Fig. 2. An example comparison of the splined tonic signal (yellow) and the re-estimated tonic signal obtained via the OSP (purple, $m$ = 1) in the real-world dataset (Electric Pulse). The reference EDA (black) and detected valleys (red circles) are also shown.

$$Y_{phasic} = Y - Y_{tonic} \quad (10)$$

Fig. 2 shows an example comparison between the splined tonic signal $\widehat{X}$ (yellow line) and the re-estimated tonic signal $Y_{tonic}$ (purple line). The main idea is to represent the tonic component as a slowly varying process that lies in a low-dimensional subspace constructed from delayed versions of the initial tonic estimate. By projecting the EDA signal onto this subspace, components that follow the slow tonic dynamics are preserved, while transient phasic fluctuations are suppressed.

### *C. Driver Estimation*

To estimate the sparse phasic driver, we used the deconvolution method using a bi-exponential function:

$$h(t) = e^{-\frac{t}{\tau_1}} - e^{-\frac{t}{\tau_2}}, \qquad t \geq 0 \quad (11)$$

where $\tau_1$ = 0.7 and $\tau_2$ = 2, as the previous research explored [18]. The kernel was normalized as $\sum_t h(t) = 1$. The convolution model is described as:

$$Y_{phasic} \approx Hp \quad (12)$$

where $H \in \mathbb{R}^{(N-m^*)\times(N-m^*)}$ is a Toeplitz convolution matrix constructed from the kernel $h(t)$, and $p \in \mathbb{R}^{N-m^*}$ is an unknown non-negative phasic driver. To estimate $p$, we resolve the non-negative least squares (NNLS) deconvolution with ridge regularization:

$$\min_{p\geq 0} \left\| Hp - Y_{phasic} \right\|_2^2 + \lambda\|p\|_2^2 \quad (13)$$

This equation can be written as:

$$\min_{p\geq 0} \frac{1}{2} p^\top Q p + f^\top p \quad (14)$$

where $Q = 2(H^\top H + \lambda I)$ and $f = -2H^\top Y_{phasic}$. The same regularization parameter $\lambda = 0.01$ is used to control smoothness. The phasic component can then be reconstructed from the estimated driver as:

$$Y_{phasic}(t) = (h * p)(t) \quad (15)$$

Finally, the phasic component re-estimated from the

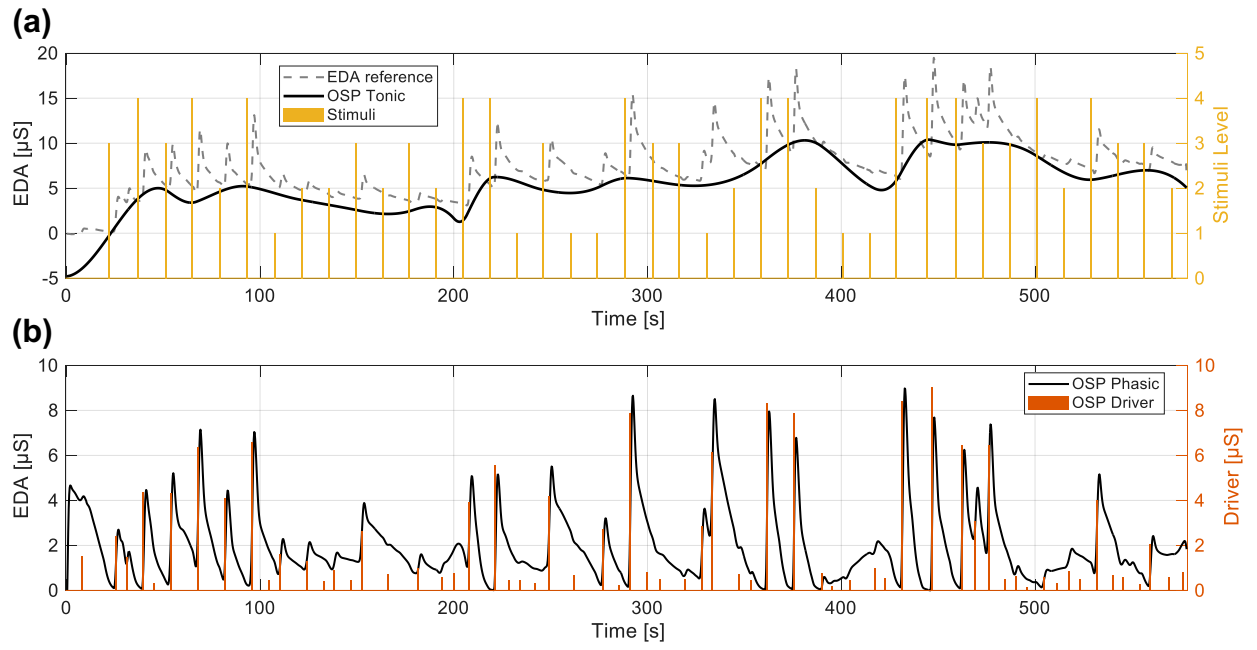

Fig. 3. An example of electrodermal activity (EDA) decomposition using the OSP method. (a) Comparison of the estimated tonic component with the reference EDA and annotated stimulus events. (b) Isolated phasic component from the OSP method and the estimated phasic driver signal derived from the phasic component.

reconstructed driver. The overall EDA signal is now defined as:

$$Y_{EDA} = Y_{phasic} + Y_{tonic} + Y_{noise} \quad (16)$$

where, $Y_{noise}$ is residual measurement noise, calculated as $Y_{EDA} - (Y_{phasic} + Y_{tonic})$. By explicitly extracting $Y_{noise}$ and utilizing NNLS, we ensure that physiological components $Y_{phasic}$ remain strictly non-negative, while allowing the noise term to absorb non-physiological noises.

To find the sparse driver, we applied two post-processing constraints: (1) amplitude thresholding = 0.01 $\mu S$ and (2) distance thresholding = 5 seconds to keep only the largest peak of the driver. Fig. 3 shows an example of EDA decomposition using the OSP method.

### *D. Evaluation Metrics*

1) Simulated EDA Dataset Evaluation: To evaluate decomposition performance using ground-truth labels, we evaluated our method in the simulated EDA dataset. For reconstruction and decomposition performance, we evaluated the root mean square error (RMSE) of the tonic and phasic components, the correlation of the phasic signal, and the coefficient of determination ($R^2$) for overall signal reconstruction. To evaluate SNA driver detection performance, we assessed the F1-score, precision, and recall. To ensure a fair comparison across all methods, an amplitude threshold of 0.01 μS (corresponding to the minimum driver amplitude) and a peak detection algorithm were applied as post-processing steps to extract discrete SCR events. To account for minor temporal shifts in the estimated drivers, we applied a ±1-second tolerance window around each ground-truth driver label. A true positive (TP) was defined as an estimated driver peak occurring within this 1-second window of a ground-truth peak. Any estimated peak outside this window was classified as a false positive (FP), and any ground-truth peak without a corresponding TP was classified as a false negative (FN).

2) Real-world Pain Database Evaluation: To compare the effects of the different stimuli, pain level 0 was defined as the 10-second baseline segment before each stimulus label, and pain levels > 0 were defined as the 10-second segment after stimulus label. This 10-second post-stimulus window is a widely accepted convention in continuous EDA pain assessment to ensure the complete physiological response is captured [26], [29]. Within each 10-second window, the maximum value of the estimated phasic driver peak was extracted as the evaluation feature. If no peak was detected, a value of zero was assigned to that segment. For the statistical analyses, this extraction yielded a perfectly balanced set of baseline (pain level = 0) and stimulus (pain level > 0) windows for each dataset: 6,960 baseline and 6,960 stimulus windows for the BioVid database; 628 baseline and 628 stimulus windows for the Electric Pulse dataset; 464 baseline and 464 stimulus windows for the Thermal Grill dataset; 60 baseline and 60 stimulus windows for the Pain-Only dataset; and 60 baseline and 60 stimulus windows for the Pain with Stroop dataset. The deconvolution kernel parameters ($\tau_1$ and $\tau_2$) were optimized in LedaLab-CDA and LedaLab-DDA. For cvxEDA and sparsEDA, we followed the parameter settings reported in the original publications. BayesianEDA and UDM estimate their model parameters automatically during execution.

3) Statistical Significance Analysis: Statistical significance of stimulus effects was assessed using a linear mixed-effects model with subject-specific random intercepts and random slopes for stimulus effect. Omnibus p-values were obtained from mixed-effects analysis of variance (ANOVA) with Satterthwaite's approximation for degrees of freedom [30]. To rigorously evaluate performance stability and address inter-subject variability, effect sizes were reported as omega squared ($\omega^2$) for the overall stimulus effect, accompanied by bootstrap-based 95% confidence intervals (CIs) computed using 1,000 iterations. We performed post-hoc analysis using estimated marginal means with Tukey-Kramer adjustment when the omnibus ANOVA was significant ($\alpha$ = 0.05).

4) Receiver Operating Characteristic (ROC) and Computational Analysis: Discrimination performance between baseline (pain level = 0) and stimulated condition (pain levels > 0) was evaluated using ROC curve analysis. For each subject, ROC curves and area under the ROC curve (AUROC) values were calculated by defining stimulated drivers (pain level > 0) as the positive class and baseline drivers (pain level = 0) as the negative class. ROC curves were obtained by varying the decision threshold across the full range of feature values and computing true positive and false positive rates at each threshold. To compare AUROC performance and elapsed computational time across the different decomposition methods, we used a linear mixed-effects model with the subject as a random factor, followed by Tukey-Kramer post-hoc comparisons (α = 0.05). To ensure robust estimation of performance stability, 95% confidence intervals for the AUROC were obtained using a 1,000-iteration subject-level bootstrap procedure. Corresponding p-values are reported in both AUROC classification and computational efficiency.

TABLE II
HEURISTIC PARAMETER OPTIMIZATION USING SIMULATED EDA DATASET (20DB SNR)

| Processing Stage | Parameter | Search area | Selected value | Optimization Metric |
|---|---|---|---|---|
| Valley Detection | Inter-valley distance (s) | [0, 30] | 20 | Phasic Correlation |
| | Prominence ($\mu S$) | [0.001, 0.1] | 0.05 | Phasic Correlation |
| | Control Point Interval (s) | [0, 30] | 20 | Phasic Correlation |
| Driver Estimation | Regularization ($\lambda$) | [0.001, 1] | 0.01 | Driver F1-Score |
| | Distance Threshold (s) | [0, 10] | 5 | Driver F1-Score |

## IV. RESULTS

### A. Heuristic parameters optimization

Table II summarizes the optimized heuristic parameters derived using a simulated EDA dataset with a 20 dB SNR. To prevent algorithmic bias, we performed a systematic grid search to identify the optimal configuration for both the valley detection and driver estimation stages. Parameters were evaluated against the simulated ground truth (with 20 dB), maximizing the phasic correlation for valley detection and the F1-score for driver estimation. These optimized parameters were then applied to evaluate five real-world datasets.

### B. Simulated EDA analysis

In the real-world dataset, there are no labels like tonic, phasic, and noise. To evaluate the EDA signal reconstruction results and SNA detection results, we compared the methods in 4 different levels (10 dB, 20 dB, 30 dB, and clean) of noise using simulated EDA signals.

Fig. 4 shows a representative example of the simulated EDA signal with a 20 dB SNR, comparing the decomposition performance of all evaluated methods against the known ground truth. As shown in Fig. 4(a), ospEDA (black) closely tracks the ground-truth tonic signal (red). In contrast, LedaLab-DDA (purple) and UDM (yellow) exhibit substantial baseline errors, while cvxEDA (light green) and BayesianEDA (light blue) display excessive fluctuations. In Fig. 4(b), the phasic component extracted by ospEDA closely aligns with the ground-truth phasic signal (dashed grey). Notably, sparsEDA produces non-physiological negative values in the phasic component, and BayesianEDA fails to resolve consecutive phasic peaks. Finally, Fig. 4(c) confirms that ospEDA closely tracks both the timing and amplitude of the discrete sympathetic phasic drivers.

Table III presents the signal decomposition performance across varying SNRs (10 dB, 20 dB, 30 dB, and clean). While cvxEDA demonstrates excellent reconstruction accuracy under clean and low-noise (30 dB) conditions, its performance rapidly degrades as noise increases. In contrast, ospEDA demonstrates superior robustness in high-noise environments. At 20 dB SNR, ospEDA achieves the lowest RMSE for both the estimated tonic (0.131) and phasic (0.132) components. At 10 dB SNR, ospEDA outperforms all other methods, maintaining the lowest phasic RMSE (0.293) with the highest phasic correlation (0.782) and $R^2$ (0.979) values. As shown in Fig. 4, LedaLab-DDA and UDM exhibit the poorest overall tonic and phasic decomposition performance, while sparsEDA and BayesianEDA consistently yield similar error magnitudes across all noise levels. Notably, LedaLab-CDA and sparsEDA maintain identical $R^2$ values regardless of the noise condition. This indicates that these algorithms do not suppress noise during processing; they strictly decompose the raw signal into tonic and phasic components.

Table IV presents the SNA detection results. LedaLab-CDA and cvxEDA achieve a recall of 1.0 across multiple noise levels but exhibit exceptionally low precision scores, especially under 30, 20, and 10 dB noise conditions, indicating a severe rate of false alarms. LedaLab-DDA and UDM show strong F1-scores and precision in clean EDA, but their performance degrades dramatically in noisy conditions. Interestingly, BayesianEDA maintains a consistent F1-score across all noise levels, which corresponds to its consistent signal reconstruction performance. While sparsEDA yields the highest precision at 10, 20, and 30 dB SNR, ospEDA consistently achieves the highest overall F1-scores across these same noise conditions. This indicates that ospEDA and sparsEDA provide the most reliable phasic driver estimation in the presence of noise artifacts.

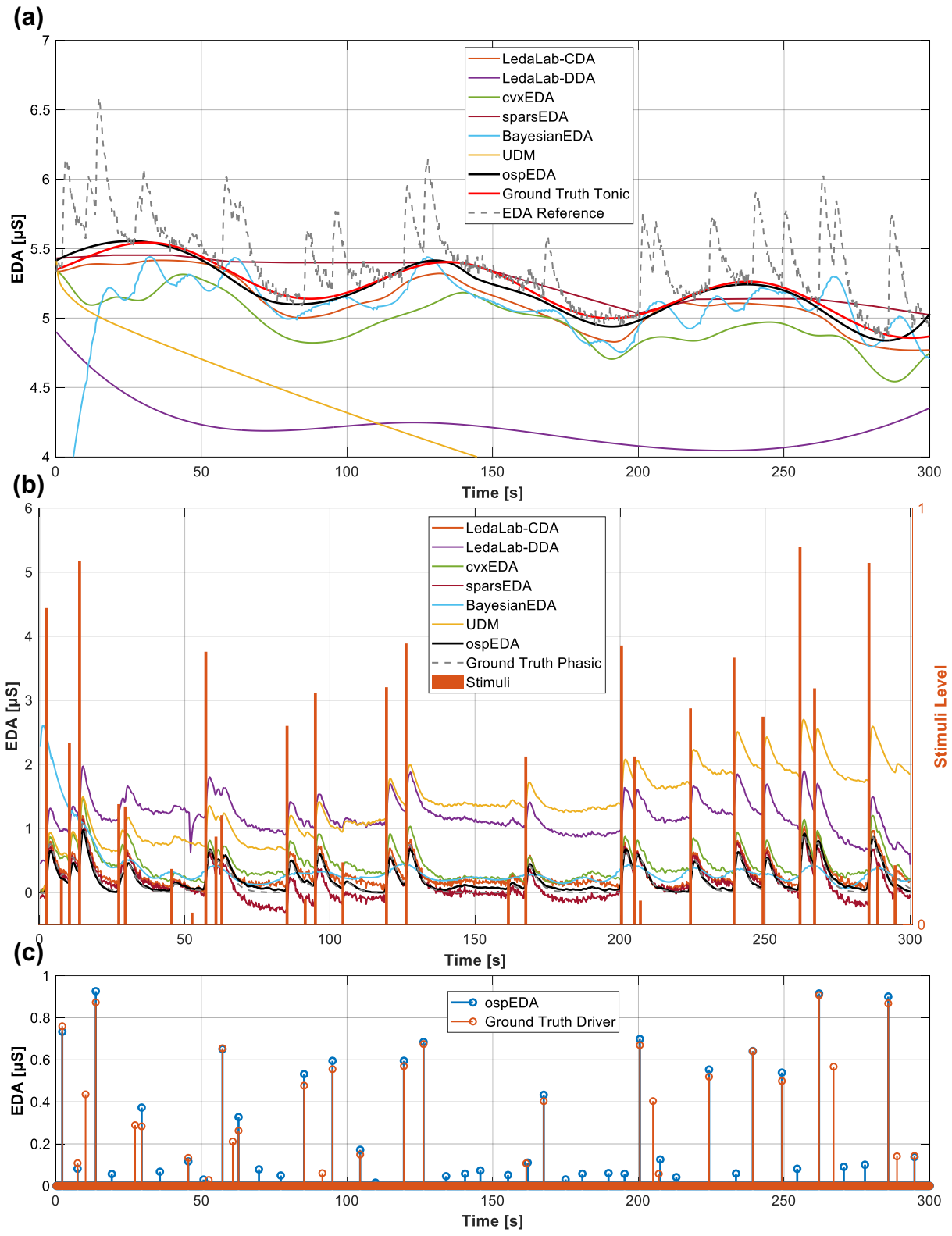

Fig. 4. An example of performance comparison in the Simulated EDA dataset (20 dB SNR). (a) Tonic estimation: ospEDA (black) tightly tracks the ground truth (red). (b) Phasic estimation: ospEDA (black) accurately recovers ground truth phasic activity (dashed grey) and strictly follows the non-negativity constraint. (c) Sparse driver estimation: ospEDA impulses (blue) precise alignment in both timing and amplitude with ground truth driver (red).

TABLE III
SIGNAL RECONSTRUCTION PERFORMANCE COMPARISON FOR SIMULATED EDA SIGNAL

| Noise Level | Metric | LedaLab-CDA | LedaLab-DDA | cvxEDA | sparsEDA | BayesianEDA | UDM | ospEDA |
|---|---|---|---|---|---|---|---|---|
| Clean | RMSE (T) | 0.054 ± 0.021 | 1.018 ± 0.231 | **0.032 ± 0.016** | 0.274 ± 0.078 | 0.350 ± 0.068 | 1.049 ± 1.083 | 0.102 ± 0.081 |
| | RMSE (P) | 0.054 ± 0.021 | 0.907 ± 0.213 | **0.034 ± 0.015** | 0.274 ±0.078 | 0.373 ± 0.063 | 1.050 ± 1.084 | 0.105 ± 0.077 |
| | Corr (P) | 0.980 ± 0.014 | 0.682 ± 0.099 | **0.994 ± 0.004** | 0.748 ± 0.120 | 0.158 ± 0.105 | 0.535 ± 0.249 | 0.935 ± 0.099 |
| | $R^2$ | **1.000 ± 0.000** | 0.053 ± 0.701 | 0.999 ± 0.000 | **1.000 ± 0.000** | 0.887 ± 0.065 | 0.991 ± 0.009 | 0.993 ± 0.004 |
| 30dB | RMSE (T) | 0.139 ± 0.035 | 1.180 ± 0.617 | **0.050 ± 0.012** | 0.277 ± 0.078 | 0.350 ± 0.069 | 1.246 ± 1.063 | 0.100 ± 0.059 |
| | RMSE (P) | 0.140 ± 0.035 | 1.171 ± 0.617 | **0.051 ± 0.012** | 0.278 ± 0.078 | 0.372 ± 0.063 | 1.246 ± 1.063 | 0.103 ± 0.063 |
| | Corr (P) | 0.968 ± 0.016 | 0.656 ± 0.159 | **0.989 ± 0.005** | 0.746 ± 0.119 | 0.158 ± 0.105 | 0.552 ± 0.235 | 0.934 ± 0.082 |
| | $R^2$ | **0.999 ± 0.000** | 0.953 ± 0.115 | **0.999 ± 0.000** | **0.999 ± 0.000** | 0.887 ± 0.065 | 0.994 ± 0.004 | 0.993 ± 0.004 |
| 20dB | RMSE (T) | 0.164 ± 0.048 | 1.503 ± 0.350 | 0.456 ± 0.130 | 0.276 ± 0.078 | 0.350 ± 0.069 | 2.701 ± 1.928 | **0.131 ± 0.058** |
| | RMSE (P) | 0.171 ± 0.047 | 1.499 ± 0.348 | 0.456 ± 0.130 | 0.280 ± 0.078 | 0.373 ± 0.063 | 2.700 ± 1.928 | **0.132 ± 0.056** |
| | Corr (P) | **0.925 ± 0.032** | 0.623 ± 0.139 | 0.872 ± 0.066 | 0.734 ± 0.116 | 0.158 ± 0.105 | 0.429 ± 0.229 | 0.912 ± 0.065 |
| | $R^2$ | 0.990 ± 0.000 | 0.981 ± 0.027 | **0.996 ± 0.000** | 0.990 ± 0.000 | 0.887 ± 0.065 | 0.995 ± 0.004 | 0.992 ± 0.004 |
| 10dB | RMSE (T) | 0.350 ± 0.106 | 3.481 ± 0.943 | 4.330 ± 1.291 | **0.278 ± 0.076** | 0.350 ± 0.069 | 2.907 ± 2.310 | 0.292 ± 0.106 |
| | RMSE (P) | 0.380 ± 0.111 | 3.475 ± 0.938 | 4.331 ± 1.291 | 0.317 ± 0.077 | 0.373 ± 0.063 | 2.906 ± 2.311 | **0.293 ± 0.104** |
| | Corr (P) | 0.735 ± 0.084 | 0.436 ± 0.146 | 0.351 ± 0.150 | 0.674 ± 0.109 | 0.157 ± 0.104 | 0.388 ± 0.223 | **0.782 ± 0.117** |
| | $R^2$ | 0.900 ± 0.004 | 0.891 ± 0.243 | 0.953 ± 0.243 | 0.900 ± 0.004 | 0.883 ± 0.065 | 0.967 ± 0.013 | **0.979 ± 0.005** |

*P = Phasic, T = Tonic, Corr = Correlation

TABLE IV
SNA DETECTION PERFORMANCE COMPARISON FOR SIMULATED EDA SIGNAL

| Noise Level | Metric | LedaLab-CDA | LedaLab-DDA | cvxEDA | sparsEDA | BayesianEDA | UDM | ospEDA |
|---|---|---|---|---|---|---|---|---|
| Clean | F1 Score | 0.618 ± 0.093 | **0.907 ± 0.118** | 0.222 ± 0.035 | 0.602 ± 0.088 | 0.354 ± 0.040 | 0.781 ± 0.157 | 0.717 ± 0.052 |
| | Precision | 0.454 ± 0.107 | **0.878 ± 0.142** | 0.126 ± 0.023 | 0.907 ± 0.077 | 0.237 ± 0.027 | 0.704 ± 0.199 | 0.708 ± 0.063 |
| | Recall | **1.000 ± 0.000** | 0.947 ± 0.100 | **1.000 ± 0.000** | 0.457 ± 0.094 | 0.701 ± 0.081 | 0.930 ± 0.075 | 0.729 ± 0.056 |
| 30dB | F1 Score | 0.158 ± 0.006 | 0.221 ± 0.024 | 0.236 ± 0.010 | 0.600 ± 0.094 | 0.354 ± 0.040 | 0.251 ± 0.145 | **0.638 ± 0.055** |
| | Precision | 0.086 ± 0.003 | 0.125 ± 0.016 | 0.134 ± 0.007 | **0.880 ± 0.089** | 0.236 ± 0.027 | 0.145 ± 0.025 | 0.568 ± 0.053 |
| | Recall | **1.000 ± 0.000** | 0.968 ± 0.083 | **1.000 ± 0.000** | 0.462 ± 0.098 | 0.701 ± 0.081 | 0.966 ± 0.036 | 0.728 ± 0.060 |
| 20dB | F1 Score | 0.168 ± 0.009 | 0.188 ± 0.028 | 0.183 ± 0.003 | 0.587 ± 0.098 | 0.356 ± 0.039 | 0.198 ± 0.019 | **0.617 ± 0.056** |
| | Precision | 0.092 ± 0.005 | 0.104 ± 0.019 | 0.101 ± 0.002 | **0.838 ± 0.126** | 0.238 ± 0.026 | 0.110 ± 0.011 | 0.542 ± 0.053 |
| | Recall | **1.000 ± 0.000** | 0.983 ± 0.039 | **1.000 ± 0.000** | 0.460 ± 0.100 | 0.704 ± 0.080 | 0.988 ± 0.024 | 0.717 ± 0.061 |
| 10dB | F1 Score | 0.156 ± 0.007 | 0.183 ± 0.026 | 0.152 ± 0.002 | 0.560 ± 0.090 | 0.357 ± 0.036 | 0.192 ± 0.030 | **0.573 ± 0.059** |
| | Precision | 0.085 ± 0.004 | 0.102 ± 0.019 | 0.082 ± 0.001 | **0.772 ± 0.146** | 0.239 ± 0.024 | 0.107 ± 0.019 | 0.498 ± 0.054 |
| | Recall | **1.000 ± 0.003** | 0.981 ± 0.033 | **1.000 ± 0.000** | 0.449 ± 0.089 | 0.706 ± 0.074 | 0.985 ± 0.039 | 0.674 ± 0.068 |

### *C. Stimulus level discrimination analysis*

Fig. 5 shows the mean and standard error of the mean (SEM) of the phasic driver across various stimulus levels for five datasets and seven decomposition methods. The y-axes are plotted on the logarithmic scale of phasic driver amplitude ($\mu S$), and x-axes display stimulus levels. Corresponding p-values for adjacent stimulus level comparisons are reported in Table I, and effect sizes ($\omega^2$) are summarized in Table VI.

Fig. 5(a) shows the results for the BioVid Heat Pain database (n = 87). LedaLab-CDA, cvxEDA, UDM, and ospEDA exhibited significant increases in phasic driver amplitude across all adjacent stimulus levels ($p < 0.05$ or $p < 0.001$). In contrast, sparsEDA and BayesianEDA showed no significant difference between level 0 and level 1 ($p \geq 0.05$), while LedaLab-DDA showed no significant difference between level 0 and level 1, as well as level 1 and level 2 ($p \geq 0.05$). As shown in Table VI, LedaLab-CDA achieved the strongest effect size ($\omega^2$ =0.394), followed by UDM (0.360), cvxEDA (0.314), and ospEDA (0.294), all of which represent large effects ($\omega^2 \geq 0.14$). While sparsEDA and BayesianEDA also demonstrate large effects, LedaLab-DDA provided only a medium effect (0.083).

Fig. 5(b) presents the results for the Electric Pulse dataset (n = 16). In this dataset, LedaLab-DDA and cvxEDA failed to reach significance in the omnibus ANOVA ($p > 0.05$); therefore, post-hoc comparisons were not performed. Among the remaining methods, only UDM showed a significant difference between level 2 and level 3 ($p = 0.048$). Despite this lack of adjacent-level sensitivity, an expected consequence of the limited number of subjects (n = 16), the overall stimulus effects were large ($\omega^2 \geq 0.14$) for all methods except for LedaLab-DDA, with BayesianEDA achieving the highest effect size (0.628), followed by sparsEDA (0.478) and ospEDA (0.458). To address the limited sample size (n = 16), we calculated 95% CIs using a 1,000-iteration subject-level bootstrap. This approach confirms that sparsEDA, BayesianEDA, UDM, and ospEDA successfully captured strong physiological responses, as the lower bounds of their respective CIs all exceed the threshold for a large effect size ($\omega^2 \geq 0.14$).

Fig. 5(c) presents the results for the Thermal Grill dataset (n = 23). In the dataset, LedaLab-DDA failed to reach significance in the omnibus ANOVA. Between levels 0 and 1, sparsEDA, BayesianEDA, UDM, and ospEDA showed significant differences ($p < 0.05$). LedaLab-CDA closely missed significance ($p = 0.053$), while cvxEDA failed to detect a significant difference ($p = 0.482$). Between levels 1 and 2, LedaLab-CDA, sparsEDA, UDM, and ospEDA showed a significant difference ($p < 0.05$). However, no method was able to detect a significant difference between levels 2 and 3 ($p \geq 0.05$). The overall stimulus effect sizes across all methods were large ($\omega^2 \geq 0.14$), except for LedaLab-DDA ($\omega^2$ = 0.113). sparsEDA achieved the highest overall effect size (0.591), followed by BayesianEDA (0.555) and ospEDA (0.547).

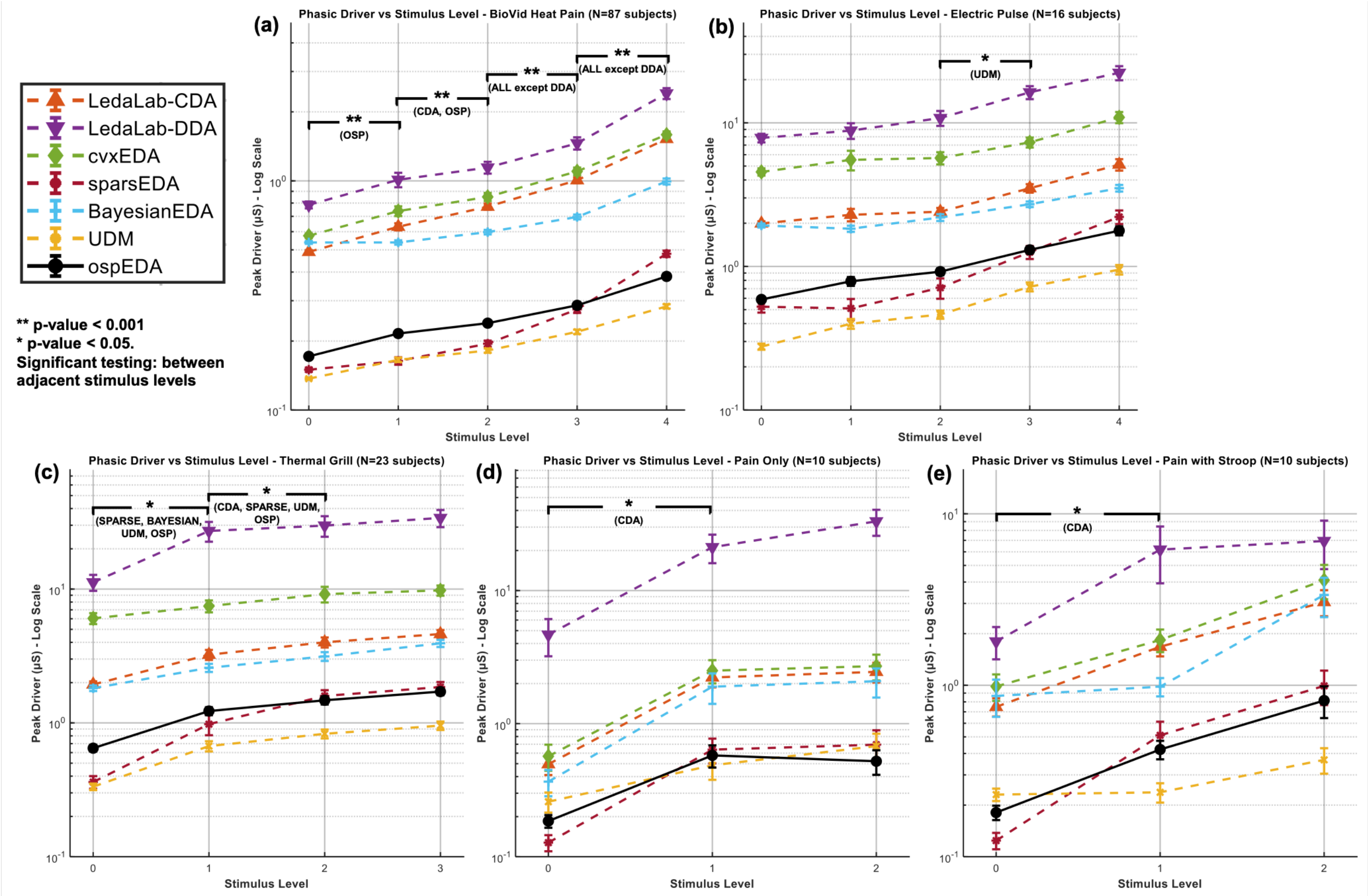

Fig. 5. Mean phasic component and phasic driver amplitude (μS) with SEM across stimulus levels for five datasets using seven EDA decomposition methods. Results are shown for (a) BioVid Heat Pain (n = 87), (b) Electric Pulse (n = 16), (c) Thermal Grill (n = 23), (d) Pain Only (n = 10), and (e) Pain with Stroop (n = 10). y-axes are plotted on a logarithmic scale, and x-axes represent stimulus levels. Statistical significance was evaluated between adjacent stimulus levels (**p < 0.001, *p < 0.05), with significant pairwise comparisons indicated above each panel.
*Abbreviations: CDA – LedaLab-CDA; DDA – LedaLab-DDA; CVX – cvxEDA; SPARSE – sparsEDA; BAYESIAN – BayesianEDA; OSP – ospEDA; ALL – CDA, DDA, CVX, SPARSE, BAYESIAN, UDM, and OSP.*

TABLE V
STIMULUS-WISE P-VALUES FOR EACH METHOD ACROSS FIVE DATASETS

| Dataset | Stimulus Levels | Methods (p-values) | | | | | | |
|---|---|---|---|---|---|---|---|---|
| | | LedaLab-CDA | LedaLab-DDA | cvxEDA | sparsEDA | BayesianEDA | UDM | ospEDA |
| BioVid | level 0 vs 1 | 0.001* | 0.171 | 0.021* | 0.563 | 1.000 | 0.002* | <0.001** |
| | level 1 vs 2 | <0.001** | 0.320 | 0.014* | 0.050* | 0.024* | 0.006* | <0.001** |
| | level 2 vs 3 | <0.001** | 0.022* | <0.001** | <0.001** | <0.001** | <0.001** | <0.001** |
| | level 3 vs 4 | <0.001** | 0.002* | <0.001** | <0.001** | <0.001** | <0.001** | <0.001** |
| Electric Pulse | level 0 vs 1 | 0.839 | – | – | 1.000 | 1.000 | 0.106 | 0.269 |
| | level 1 vs 2 | 0.999 | – | – | 0.891 | 0.401 | 0.900 | 0.719 |
| | level 2 vs 3 | 0.152 | – | – | 0.090 | 0.303 | 0.048* | 0.089 |
| | level 3 vs 4 | 0.473 | – | – | 0.465 | 0.127 | 0.524 | 0.413 |
| Thermal Grill | level 0 vs 1 | 0.053 | – | 0.482 | 0.022* | 0.024* | 0.004* | 0.002* |
| | level 1 vs 2 | 0.021* | – | 0.332 | 0.039* | 0.109 | 0.041* | 0.015* |
| | level 2 vs 3 | 0.696 | – | 0.962 | 0.975 | 0.428 | 0.660 | 0.286 |
| Pain Only | level 0 vs 1 | 0.029* | – | 0.078 | – | 0.287 | – | 0.109 |
| | level 1 vs 2 | 0.193 | – | 0.402 | – | 0.737 | – | 0.791 |
| Pain with Stroop | level 0 vs 1 | 0.026* | – | – | – | – | – | – |
| | level 1 vs 2 | 0.096 | – | – | – | – | – | – |

Note: Statistical significance is denoted as follows: p < 0.05 (*), p < 0.001 (**). Empty cells (–) indicate that post-hoc comparisons were not performed due to non-significant main effects (p ≥ 0.05).

Furthermore, LedaLab-CDA, sparsEDA, BayesianEDA, UDM, and ospEDA maintained large effect sizes even at the lower bounds of their 95% CIs ($\omega^2 \geq 0.14$).

Fig. 5(d) presents the results for the Pain Only dataset (n = 10). In this dataset, LedaLab-DDA, sparsEDA, and UDM failed to reach significance in the omnibus ANOVA. Among the remaining methods, only LedaLab-CDA showed a significant difference between levels 0 and 1 ($p = 0.029$). However, all methods demonstrated large stimulus effect sizes ($\omega^2 \geq 0.14$). LedaLab-CDA achieved the strongest effect size (0.417), followed by BayesianEDA (0.352) and UDM (0.346). To address the limited sample size, we validated the effect sizes using a 1,000-iteration subject-level bootstrap, confirming that LedaLab-CDA, cvxEDA, and BayesianEDA consistently

TABLE VI
EFFECT SIZES ($\omega^2$) OF EDA DECOMPOSITION METHODS ACROSS DATASETS

| Dataset | Methods (effect size ($\omega^2$)) | | | | | | |
|---|---|---|---|---|---|---|---|
| | LedaLab-CDA | LedaLab-DDA | cvxEDA | sparsEDA | BayesianEDA | UDM | ospEDA |
| BioVid | **0.394 [0.333, 0.525]** ** | 0.083 [0.042, 0.246] * | 0.314 [0.224, 0.414] ** | 0.270 [0.207, 0.366] ** | 0.261 [0.185, 0.448] ** | 0.360 [0.267, 0.457] ** | 0.294 [0.238, 0.458] ** |
| Electric Pulse | 0.282 [0.074, 0.754] ** | 0.109 [0.014, 0.340] * | 0.178 [0.069, 0.599] ** | 0.478 [0.228, 0.818] ** | **0.628 [0.450, 0.823]** ** | 0.297 [0.175, 0.693] ** | 0.458 [0.227, 0.806] ** |
| Thermal Grill | 0.294 [0.192, 0.660] ** | 0.113 [0.017, 0.274] * | 0.379 [0.109, 0.566] ** | **0.591 [0.423, 0.734]** ** | 0.555 [0.428, 0.711] ** | 0.475 [0.221, 0.605] ** | 0.547 [0.328, 0.728] ** |
| Pain Only | **0.417 [0.274, 0.713]** ** | 0.237 [0.028, 0.574] ** | 0.297 [0.171, 0.598] ** | 0.149 [0.024, 0.535] ** | 0.352 [0.159, 0.730] ** | 0.346 [0.046, 0.435] ** | 0.292 [0.116, 0.612] ** |
| Pain with Stroop | **0.338 [0.180, 0.719]** ** | 0.045 [0.000, 0.554] | 0.209 [0.100, 0.559] ** | 0.190 [0.082, 0.518] ** | 0.065 [0.005, 0.567] * | 0.000 [0.000, 0.448] | 0.175 [0.064, 0.536] ** |
| Range of $\omega^2$ | 0.282 – 0.417 | 0.045 – 0.237 | 0.178 – 0.379 | 0.149 – 0.591 | 0.065 – 0.628 | 0.000 – 0.475 | 0.175 – 0.547 |

Note: Effect sizes are reported as $\omega^2$ [95% CI] derived from 1,000 subject-level bootstrap iterations. Bold indicates the largest effect size. Asterisks denote the magnitude of the effect: * indicates a medium effect ($\omega^2 \geq 0.06$), and ** indicates a large effect ($\omega^2 \geq 0.14$)

maintained large effect sizes even at the lower bounds of their 95% CIs ($\omega^2 \geq 0.14$).

Fig. 5(e) presents the results for the Pain with Stroop dataset (n = 10). The combination of a highly limited sample size and the confounding cognitive load of the Stroop task severely impacted discrimination performance across most algorithms. Only LedaLab-CDA was able to detect a significant difference between levels 0 and 1 (p = 0.026). However, LedaLab-CDA, cvxEDA, sparsEDA, and ospEDA all demonstrated large overall stimulus effect sizes. LedaLab-CDA achieved the strongest effect size (0.338), followed by cvxEDA (0.209), sparsEDA (0.190), and ospEDA (0.175). To address the limited discrimination performance, we validated the effect sizes using a 1,000-iteration subject-level bootstrap. This confirmed that LedaLab-CDA consistently maintained a large effect size even at the lower bound of their 95% CIs ($\omega^2 \geq 0.14$). whereas cvxEDA, sparsEDA, and ospEDA maintained medium effect sizes at their respective lower bounds ($\omega^2 \geq 0.06$).

Overall, across all five real-world pain datasets, LedaLab-CDA demonstrated the strongest stimulus discrimination performance. It achieved the highest effect sizes in the BioVid, Pain Only, and Pain with Stroop datasets, maintaining a consistently high and narrow range of large effect sizes ($\omega^2$ = 0.282 – 0.417). While algorithms such as sparsEDA and BayesianEDA achieved the highest performances in Thermal Grill and Electric Pulse datasets, their sensitivity fluctuated considerably depending on the dataset (e.g., dropping effect size in the Pain Only and Pain with Stroop datasets). Notably, the proposed ospEDA method demonstrated cross-dataset reliability. Alongside LedaLab-CDA and cvxEDA, ospEDA successfully extracted large overall stimulus effect sizes ($\omega^2 \geq$ 0.14) across every single dataset evaluated, with its effect magnitudes ranging from 0.175 to 0.547. This consistent extraction of large effect sizes highlights that while LedaLab-CDA provides the highest stimulus level discrimination, ospEDA showed relatively consistent performance across the evaluated experimental conditions.

### D. Non-stimulus vs. Stimulus Classification Performance

Fig. 6 shows the overall ROC curves and AUROC distributions across the seven EDA decomposition methods for binary stimulus classification between baseline (pain level = 0) and stimulated conditions (pain level > 0). Across the aggregated five datasets (n = 146), ospEDA achieved the

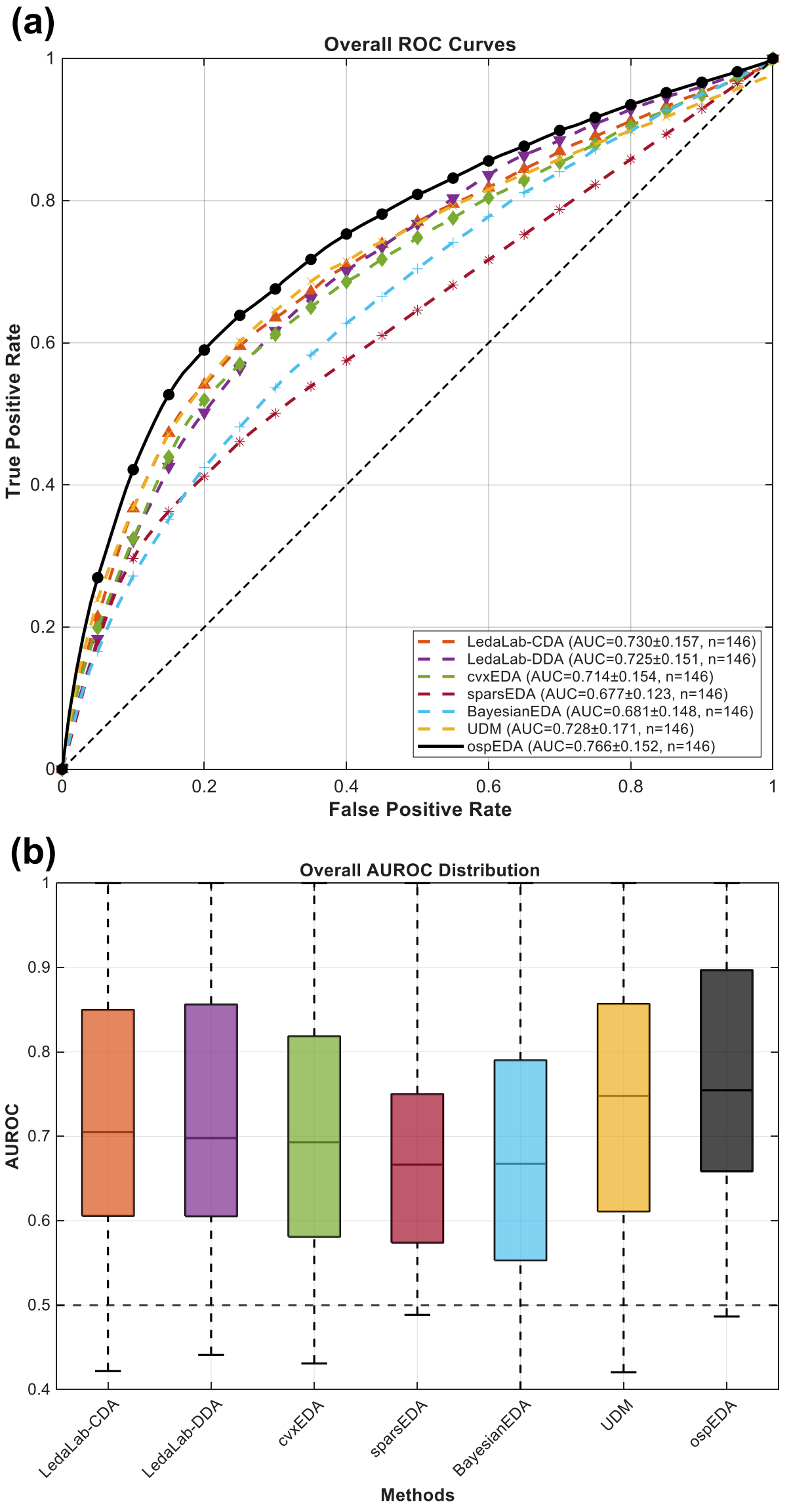

Fig. 6. Overall binary classification performance of the seven EDA decomposition methods across all aggregated datasets (n = 146). (a) ROC curves comparing true positive and false positive rates. (b) Boxplots illustrating the distribution of AUROC scores.

TABLE VII
AUROC RESULTS OF EDA DECOMPOSITION METHODS ACROSS FIVE REAL-WORLD DATASETS

| Method | Dataset (Mean ± SD [95% CI], n = 146 total) | | | | | | | | |
|---|---|---|---|---|---|---|---|---|---|
| | BioVid (n = 87) | Electric Pulse (n = 16) | Thermal Grill (n = 23) | Pain Only (n = 10) | Pain with Stroop (n = 10) | Mean ± SD | Q1 | Median | Q3 |
| LedaLab-CDA | 0.690 ± 0.124* [0.663, 0.717] | 0.663 ± 0.133 [0.598, 0.723] | 0.758 ± 0.148* [0.698, 0.814] | 0.928 ± 0.173* [0.817, 1.000] | **0.925 ± 0.165 [0.825, 1.000]** | 0.730 ± 0.157 | 0.606 | 0.705 | 0.850 |
| LedaLab-DDA | 0.689 ± 0.120* [0.663, 0.713] | 0.663 ± 0.147 [0.595, 0.729] | 0.733 ± 0.141 [0.681, 0.787] | 0.915 ± 0.145* [0.819, 0.981] | **0.933 ± 0.120 [0.853, 0.994]** | 0.725 ± 0.151 | 0.605 | 0.698 | 0.856 |
| cvxEDA | 0.663 ± 0.116 [0.638, 0.688] | 0.665 ± 0.131 [0.604, 0.725] | 0.747 ± 0.148* [0.687, 0.802] | **0.944 ± 0.111 [0.872, 0.992]** | **0.922 ± 0.146 [0.830, 0.989]** | 0.714 ± 0.154 | 0.581 | 0.693 | 0.818 |
| sparsEDA | 0.635 ± 0.101 [0.614, 0.655] | 0.643 ± 0.075 [0.606, 0.678] | 0.728 ± 0.096 [0.690, 0.767] | 0.818 ± 0.151 [0.725, 0.901] | **0.835 ± 0.121 [0.765, 0.901]** | 0.677 ± 0.123 | 0.574 | 0.667 | 0.750 |
| BayesianEDA | 0.643 ± 0.126 [0.618, 0.669] | 0.656 ± 0.124 [0.599, 0.716] | 0.733 ± 0.100 [0.694, 0.776] | 0.867 ± 0.160* [0.767, 0.950] | 0.747 ± 0.245* [0.597, 0.886] | 0.681 ± 0.148 | 0.553 | 0.667 | 0.790 |
| UDM | 0.706 ± 0.131* [0.679, 0.732] | **0.777 ± 0.131 [0.710, 0.835]** | **0.815 ± 0.137 [0.758, 0.866]** | 0.793 ± 0.167 [0.687, 0.889] | 0.576 ± 0.342 [0.342, 0.792] | 0.728 ± 0.171 | 0.611 | 0.748 | 0.857 |
| ospEDA | **0.717 ± 0.133 [0.689, 0.744]** | 0.746 ± 0.137* [0.679, 0.806] | **0.819 ± 0.122 [0.770, 0.866]** | 0.931 ± 0.148* [0.830, 0.994] | **0.931 ± 0.150 [0.833, 1.000]** | 0.766 ± 0.152 | 0.659 | 0.755 | 0.897 |

Note: AUROCs are reported as Mean ± SD [95% CI] derived from 1,000 subject-level bootstrap iterations. Bold fonts indicate the best-performing method in each column, which is significantly higher than the others ($p < .05$). Methods marked with * are not significantly different from the boldfaced method.

highest overall mean AUROC performance (0.766 ± 0.152), as shown in Fig. 6(a). This was followed by LedaLab-CDA (0.730 ), UDM (0.728 ± 0.171), and LedaLab-DDA (0.725 ± 151). In contrast, BayesianEDA and sparsEDA demonstrated the lowest overall classification performance (0.681 ± 0.148 and 0.677 ± 0.123, respectively). Furthermore, the AUROC distribution boxplots (Fig. 6(b)) show that ospEDA not only achieved the highest performance but also the greatest consistency across subjects, providing the highest first quartile (Q1 = 0.659), median (0.765), and third quartile (Q3 = 0.897).

Table VII presents the AUROC results of EDA decomposition methods across the five real-world datasets. For the BioVid Heat Pain dataset (n = 87), ospEDA achieved the highest classification performance with an AUROC of 0.717 ± 0.133 (95% CIs [0.689, 0.744]). UDM (0.706), LedaLab-CDA (0.690), and LedaLab-DDA (0.689) demonstrated comparable performance with no significant difference from ospEDA. For the Electric Pulse dataset (n = 16), UDM achieved the highest classification performance with an AUROC of 0.777 ± 0.131 (95% CIs [0.710, 0.835]), while ospEDA exhibited comparable performance (0.746). For the Thermal Grill dataset (n = 23), ospEDA achieved the highest AUROC of 0.819 ± 0.122 (95% CI [0.770, 0.866]). UDM (0.815), LedaLab-CDA (0.758), and cvxEDA (0.747) also exhibited statistically comparable performances. For the Pain Only dataset (n = 10), cvxEDA achieved the highest AUROC of 0.944 ± 0.111 (95% CI [0.872, 0.992]). However, ospEDA (0.931), LedaLab-CDA (0.928), and LedaLab-DDA (0.915) provided statistically comparable performances. For the Pain with Stroop dataset, LedaLab-DDA achieved the best performance with an AUROC of 0.933 ± 0.120 (95% CI [0.853, 0.994]), closely followed by ospEDA (0.931), LedaLab-CDA (0.925), cvxEDA (0.922), and sparsEDA (0.835). In contrast, UDM exhibited a substantial decline in performance, with an AUROC of 0.576.

Overall, for binary non-stimulus versus stimulus classification, ospEDA achieved the highest mean AUROC across the aggregated datasets (0.766 ± 0.152). However, its performance varied across datasets: ospEDA achieved the best performance in the BioVid and Thermal Grill datasets, whereas in the remaining datasets, its AUROC was statistically comparable to that of competing methods. Specifically, in the Electric Pulse, Pain Only, and Pain with Stroop datasets, UDM (0.777), cvxEDA (0.944), and LedaLab-DDA (0.933) achieved the highest AUROC, respectively. These results suggest that ospEDA provides strong and relatively consistent classification performance across diverse experimental conditions.

### *E. Computational Time*

Table VIII summarizes the computational time results across the five datasets. The BioVid dataset contained the longest average signal lengths (6168 ± 169.1 samples; approximately 25.7 minutes), while the Pain Only dataset contained the shortest (1538 ± 152.0 samples; approximately 6.4 minutes). Across all datasets, sparsEDA was consistently the most computationally efficient, requiring less than 0.1 seconds to process even the longest signals. Statistical analysis confirmed that sparsEDA, cvxEDA, and LedaLab-CDA yielded the fastest computation times overall. LedaLab-DDA and ospEDA ranked next in efficiency; however, both methods exhibited noticeably longer computation times in the BioVid database. Finally, BayesianEDA and UDM showed the least computational

TABLE VIII
COMPUTATION TIME (SEC) RESULTS ACROSS DATASETS

| Method | Dataset (Sample Length) | | | | |
|---|---|---|---|---|---|
| | BioVid (6168 ± 169.1) | Electric Pulse (2397 ± 284.5) | Thermal Grill (1709 ± 341.7) | Pain Only (1538 ± 152.0) | Pain with Stroop (1691 ± 113.7) |
| LedaLab-CDA | **3.417 ± 0.906** | **1.007 ± 0.325** | **0.731 ± 0.265** | **0.580 ± 0.404** | **0.636 ± 0.388** |
| LedaLab-DDA | 18.48 ± 5.756 | **3.417 ± 1.773** | **2.048 ± 0.948** | **1.732 ± 0.477** | **1.581 ± 0.400** |
| cvxEDA | **0.440 ± 0.125** | **0.202 ± 0.139** | **0.141 ± 0.063** | **0.113 ± 0.042** | **0.160 ± 0.154** |
| sparsEDA | **0.058 ± 0.030** | **0.057 ± 0.026** | **0.042 ± 0.019** | **0.031 ± 0.019** | **0.037 ± 0.024** |
| BayesianEDA | 74.65 ± 13.78 | 161.3 ± 177.8 | 187.1 ± 166.1 | 20.27 ± 3.385 | 21.29 ± 1.604 |
| UDM | 104.1 ± 48.16 | **21.32 ± 11.25** | **13.28 ± 8.704** | 10.29 ± 6.250 | 10.80 ± 4.763 |
| ospEDA | 18.82 ± 1.630 | **1.708 ± 0.594** | **0.835 ± 0.619** | **0.561 ± 0.153** | **0.695 ± 0.158** |

Bold fonts indicate the best-performing method in each column, which is significantly faster than the others ($p < .05$).

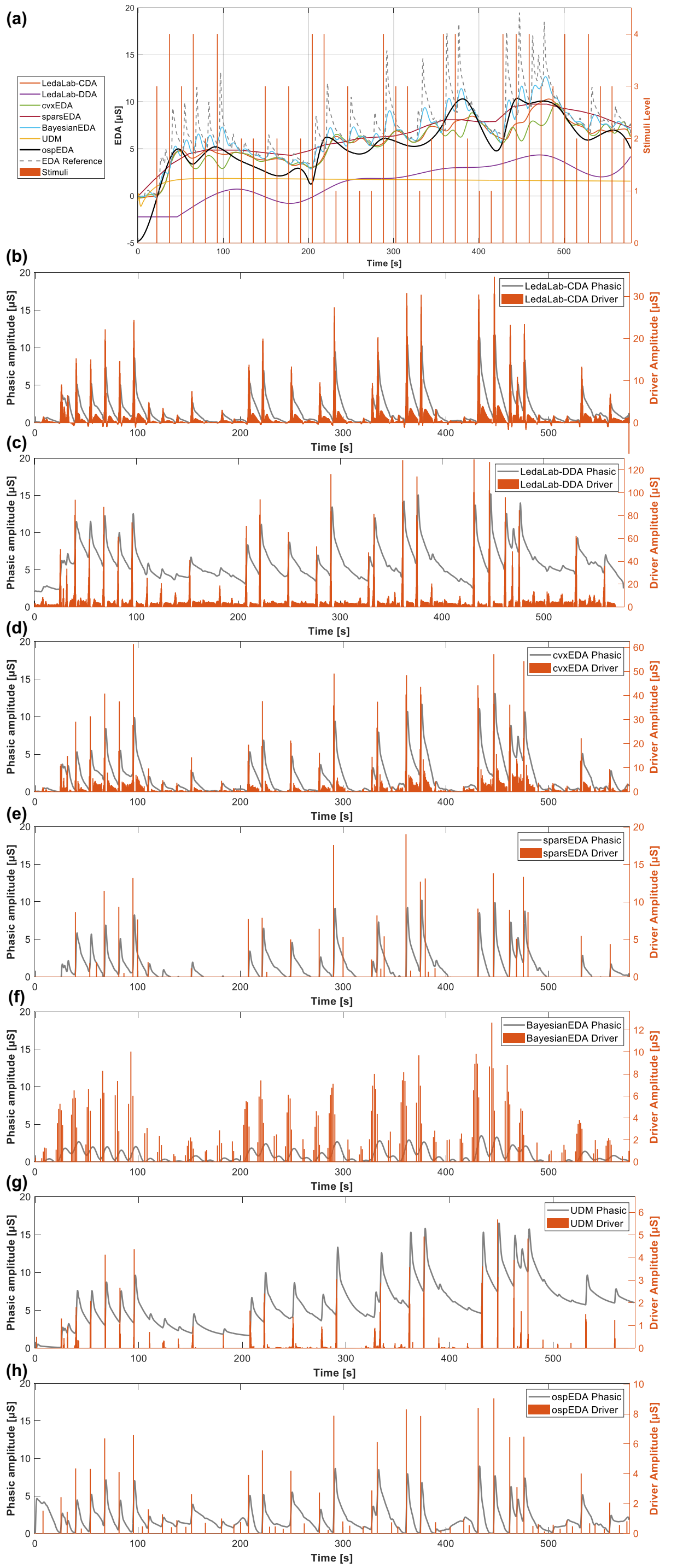

Fig. 7. An example comparison of seven EDA decomposition methods showing tonic, phasic, and driver estimation in the Electric Pulse Dataset. (a) Tonic separation in EDA signals across all methods. Phasic and driver estimation for: (b) Ledalab-CDA, (c) Ledalab-DDA, (d) cvxEDA, (e) sparsEDA, (f) BayesianEDA, (g) UDM, and (h) ospEDA ($m = 1$).

efficiency.

### *F. Visual Comparison of Driver Estimation*

Fig. 7 provides a comparison of the seven decomposition methods, including the tonic, phasic, and driver estimations for a sample recording from the Electric Pulse dataset. For the tonic component (Fig. 7(a)), LedaLab-DDA (purple) and UDM (yellow) exhibited inaccurate tonic separation, which failed to capture the underlying slow shifts of the tonic component. In contrast, BayesianEDA (blue) exhibited tonic estimates that tracked phasic fluctuations. In the phasic and driver domains, LedaLab-CDA (Fig. 7(b)), LedaLab-DDA (Fig. 7(c)), and cvxEDA (Fig. 7(d)) produced numerous phasic driver responses, increasing the likelihood of false SNA detections. BayesianEDA (Fig. 7(f)) also produced numerous spurious drivers, although they exhibited a clearer amplitude difference between the non-stimulus and stimulus periods. sparsEDA (Fig. 7(e)) generated the sparsest driver representation; while this effectively reduced false alarms, it resulted in many missed physiological SNA responses (e.g., the lack of drivers between 120 and 200 seconds). Both UDM (Fig. 7(g)) and ospEDA (Fig. 7(h)) produced physiologically plausible, sparse drivers. However, UDM frequently detected multiple SNAs in short succession, which violates the established absolute refractory period of sudomotor nerve activity. In contrast, ospEDA showed clearer separation of the tonic and phasic components while preserving plausible consecutive impulses. Overall, Fig. 7 demonstrates that while most methods can separate tonic and phasic components, they differ considerably in stability and physiological validity.

## V. Discussion

This study proposed a novel EDA decomposition method using an orthogonal subspace projection (OSP) approach. Our evaluations across five real-world datasets and a simulated EDA dataset demonstrate that ospEDA provides generally strong and consistent performance for EDA decomposition and stimulus classification. Specifically, this research makes three key contributions: (1) it demonstrates precise tonic and phasic estimation, alongside accurate driver timing estimation, validated on a ground-truth simulated EDA dataset; (2) it evaluates robust stimulus discrimination and binary classification performance across five diverse real-world datasets, supported by rigorous statistical validation; and (3) by utilizing five different datasets, it provides evidence of the method's stability across diverse experimental settings, which have not been previously validated in the EDA decomposition literature [16]–[21].

While no single method achieved absolute superiority across every metric, our evaluations highlighted distinct trade-offs among the current methods. For example, LedaLab-CDA demonstrated strong effect sizes for stimulus discrimination; however, simulation results indicate a lack of driver sparsity and an absence of noise reduction mechanisms. Bayesian approaches such as BayesianEDA and UDM show high computational times. LedaLab-DDA and cvxEDA work very well under noiseless conditions, while sparsEDA produces sparse drivers but occasionally violates the non-negative phasic physiological constraint. In light of these trade-offs, ospEDA provides a competitive and relatively stable framework across diverse noise conditions.

It achieved the best RMSE, phasic correlation, and $R^2$ scores in the 10 dB noise simulated EDA dataset, the highest F1-scores for SNA detection across 10, 20, and 30 dB SNRs, and the best binary classification AUROC with large stimulus effect sizes ($\omega^2 \geq 0.14$) across the five real-world datasets. The main idea underlying the stability of ospEDA is the orthogonal subspace projection. By modeling the tonic component as a slowly varying signal within a low-dimensional subspace, ospEDA effectively suppresses noise and separates the tonic and phasic components. This clean decomposition enables the subsequent non-negative least squares model to extract accurate phasic drivers. Furthermore, by objectively optimizing the algorithm's heuristic parameters using a simulated EDA dataset, the framework avoids overfitting. This optimization helps the model capture physiologically plausible consecutive sympathetic responses.

However, this study has limitations. First, the algorithm exhibits quadratic computational complexity with respect to the signal length since it constructs a two-dimensional orthogonal subspace from the entire signal. This can result in substantial computational time when applied to long recordings (e.g., continuous monitoring sessions exceeding 6,000 samples (approximately 25 minutes at 4 Hz), such as the BioVid dataset). To mitigate this numerical instability, we computed $V_m(V_m^\top V_m)^{-1}V_m^\top$ instead of $V_m^\top V_m$, addressing ill-conditioning through reciprocal condition number checks. However, this strategy does not fully resolve the high computational burden for long signals. A window-based OSP approach could be adopted to enable real-time applications.

Second, it is important to acknowledge that the Pain Only and Pain with Stroop datasets include a relatively small number of subjects (n = 10) and a limited number of trials (60 baseline and 60 stimulus). While our use of a linear mixed-effects model with 1,000-iteration bootstrapping helps mitigate the impact of individual variability, these smaller cohorts may limit the generalizability of the results.

Third, the ospEDA exhibits boundary instability at the beginning and end of the EDA recordings, as shown in Fig. 7. Because the initial tonic estimate relies on spline interpolation, the lack of surrounding control points at the signal edges makes the decomposition inherently less stable. Furthermore, the sample delay operator used to construct the orthogonal subspace can exacerbate these edge effects. While ospEDA achieves robust performance across diverse noise conditions, these boundary estimation errors remain a limitation.

Finally, a fundamental challenge remains in real-world EDA research: the ground-truth tonic and phasic components are inherently unknown. To address this limitation, our study introduced a simulated EDA dataset with established ground-truth tonic and phasic components. While this allowed for an objective evaluation of decomposition performance across varying SNRs, it is important to acknowledge that this simulated data is algorithmically generated. Consequently, it may not fully capture the complex noise profiles, artifact dynamics, and inter-subject variability present in real-world physiological recordings. Future research should continue to bridge this gap by developing more sophisticated, physiologically modeled simulations alongside expanded real-world validation.

## VI. Conclusion

In this study, we introduced ospEDA, a novel framework for the decomposition of electrodermal activity using orthogonal subspace projection. To overcome the limitations of prior single-dataset evaluation and address challenges such as noise sensitivity, we evaluated ospEDA across a simulated dataset (under noiseless, 30 dB, 20 dB, and 10 dB noise conditions) and five diverse real-world pain datasets.

Under noisy conditions, ospEDA demonstrated highly robust reconstruction performance. It achieved the best tonic (0.131) and phasic (0.132) root mean square errors (RMSE) under 20 dB noise, as well as the best phasic RMSE (0.293), phasic correlation (0.782), $R^2$ score (0.979) under severe 10dB noise. Furthermore, ospEDA achieved the highest F1-scores (0.638, 0.617, and 0.573) across the 30 dB, 20 dB, and 10 dB noise conditions, respectively, confirming its precision in sparse driver timing estimation.

In the real-world datasets evaluations, ospEDA achieved the highest overall binary classification performance (mean AUROC = 0.766) and consistently demonstrated large effect sizes across all five datasets. These results suggest that ospEDA provides strong overall classification performance and relatively consistent effect sizes across the evaluated experimental settings. Overall, ospEDA represents a promising decomposition framework for continuous EDA analysis under noisy conditions. Given the broad clinical and research applications of EDA, these results suggest that ospEDA may be useful for future real-time and domain-specific implementations.

## VII. Code availability

The ospEDA code and its implementation for simulated signals are available at: https://github.com/yongbin98/ospEDA

Acknowledgements: This work was supported by DHA23B-003